\documentclass{cernyrep}
\usepackage{amsmath}
\usepackage{siunitx}

\usepackage[T1]{fontenc}
\usepackage[bookmarks, colorlinks=true, linktoc=page, pdftex, linkcolor=black, citecolor=black, urlcolor=blue]{hyperref}

\usepackage{fancyhdr}
\fancyhfoffset{4 mm}
\fancypagestyle{ARTTITLE}{%
\fancyhf{}
\lhead{\raggedright {\it Mechanical \& Materials Engineering for Particle Accelerators and Detectors}\raggedright \\ CERN Accelerator School Proceedings ---  Sint-Michielsgestel,  Netherlands, 2024\hfill}
\lfoot{\hspace{3mm} Available online at \url{https://cas.web.cern.ch/previous-schools}}
\rfoot{\thepage\hspace*{3mm}}

}

\usepackage{graphicx}
\usepackage{booktabs}
\usepackage{siunitx}
\usepackage{float}

\begin{document}

\title{Beam Intercepting Devices}
\author{Davide Reggiani}
\institute{Paul Scherrer Institute (PSI), Villigen, Switzerland}

\begin{abstract}
Beam Intercepting Devices (BIDs) include targets, scrapers, collimators, protection absorbers and beam dumps.
They enable secondary particle production, shape or clean beams, and protect sensitive components by concentrating beam losses into shielded locations.
In high-power proton machines, BIDs operate close to thermo-mechanical limits under intense radiation fields, and their reliability directly impacts accelerator availability.
This review summarizes the dominant design drivers (energy deposition, temperature gradients, thermal stress, fatigue, radiation damage and activation), outlines a pragmatic design workflow combining energy-deposition assessment with coupled thermal/structural and fluid dynamic analyses, and reviews representative BIDs at PSI's High Intensity Proton Accelerator (HIPA), including current hardware and developments for the IMPACT project (Isotope and Muon Production using Advanced Cyclotron and Target technology).
\end{abstract}


\maketitle
\thispagestyle{ARTTITLE}

\section{Introduction}
Beam intercepting devices (BIDs) are the components in an accelerator complex that are intentionally exposed to the beam.
They enable secondary particle production (targets), beam shaping and cleaning (scrapers, collimators, slits), machine protection (protection collimators, absorbers), and safe disposal of unused beam (beam dumps) \cite{bib:CERN-Calviani}.
Because BIDs localize losses, they are also central to maintainability: concentrating beam losses into shielded locations avoids spreading activation along extended sections of beam line.

In high-intensity proton facilities, the combination of high beam power, small spot sizes and stringent availability goals makes BID design a multi-disciplinary optimization problem.
Design decisions are constrained simultaneously by particle-physics requirements, thermo-mechanical limits, cooling performance, radiation damage and activation, manufacturing feasibility, and remote-handling logistics.
This lecture reviews these key design aspects and illustrates them through case studies from PSI-HIPA accelerator complex~\cite{bib:HIPA2021,bib:HIPAStatus2021}.

\section{Design Drivers and Dominant Failure Modes}
\subsection{Beam Time Structure and Energy Deposition}
The thermo-mechanical response of a BID depends strongly on how power is deposited in time.
Continuous-wave (CW) beams deposit power quasi-steadily and often justify steady-state thermal assumptions.
Pulsed beams deposit energy in bursts with high instantaneous power, increasing the relevance of transient temperature fields, stress waves, and low-cycle fatigue.
For circular machines, large stored beam energy is disposed on dumps, and design must withstand extreme short-duration loads~\cite{bib:LHC-BD,bib:CERN2022}.

The primary output required for subsequent design steps is the spatial distribution of deposited power (or energy per pulse).
Analytical estimates may be suitable for simplified geometries, whereas realistic devices typically require Monte Carlo particle transport calculations~\cite{bib:FLUKA-WEB,bib:FLUKA-Paper,bib:MCNP-Paper}, which produce volumetric heat-source maps for thermal analysis.

\subsection{Thermal Limits and Hot-Spot Control}
High-power BIDs can experience extreme temperatures (hundreds to thousands of \si{\celsius}) and, more critically, large spatial gradients.
Hot spots and steep gradients are often the root cause of high thermal stress and the trigger for plastic deformation or cracking.
Cooling concepts must therefore be designed according to the beam power distribution map and integrated within the BID's geometry.

Heat dissipation pathways include conduction within the solid, convection to cooling fluids (often water), and thermal radiation between surfaces.
Cooling design must also address boiling risk, cavitation, erosion/corrosion, pressure drop, and flow stability.

\subsection{Stress, Deformation, and Fatigue}
Thermal fields drive stress and deformation through constrained thermal expansion.
In many BIDs, the~primary stress driver is non-uniform temperature distribution generated by beam interaction.
If stress exceeds yield strength, plastic strain accumulates; if it exceeds ultimate tensile strength, fracture is likely.

\subsection{Radiation Damage and Activation}
High radiation fields cause displacement damage (DPA) and activation, affecting material properties and turning remote handling into an essential technology.
These constraints influence segmentation, interface design, and shielding.

\section{BID Design Workflow}

\begin{figure}[b!]
\centering
\includegraphics[width=0.99\linewidth]{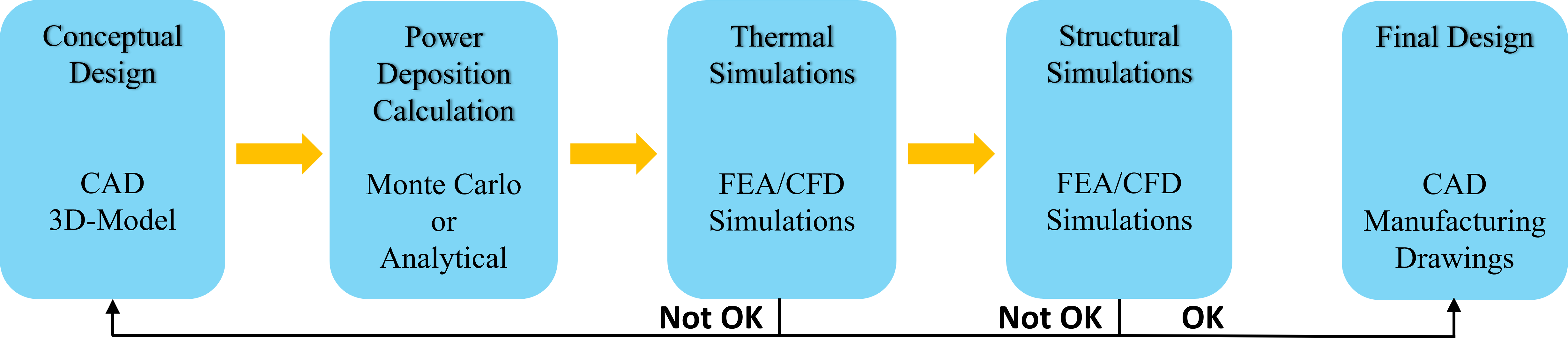}
\caption{Diagram representation of the BID design workflow.}
\label{fig:Workflow}
\end{figure}

BID design combines beam physics, thermal engineering, structural mechanics, materials science, and cooling system design. Key choices include geometry, material, power-deposition management, steady-state and transient thermal response, stress and fatigue, cooling performance, vacuum and shielding integration, lifetime strategy, manufacturability, installation and removal, and diagnostics. In practice, the~design process is iterative: conceptual geometry is first defined, energy deposition is then computed analytically or with Monte Carlo tools, and the results are transferred to thermal and structural Finite Element Analysis (FEA) and/or Computational Fluid Dynamics (CFD) simulations. If temperatures or stresses exceed allowable limits, the design is revised and reanalysed until acceptable margins are obtained. This design workflow is represented in Fig.~\ref{fig:Workflow}. Modern tools such as FLUKA~\cite{bib:FLUKA-WEB,bib:FLUKA-Paper} or MCNP~\cite{bib:MCNP-Paper} for particle transport and Ansys\textsuperscript{\textregistered}~\cite{bib:Ansys} for coupled thermo-mechanical analysis are indispensable, but the~computational cost can be substantial. For example, a quarter-symmetry FEA/CFD simulation of the~PSI SINQ spallation target with more than one million cells required about two months on a 20-core machine with \SI{1.5}{TB} RAM.

\begin{figure}[b!]
\centering
\includegraphics[width=\linewidth]{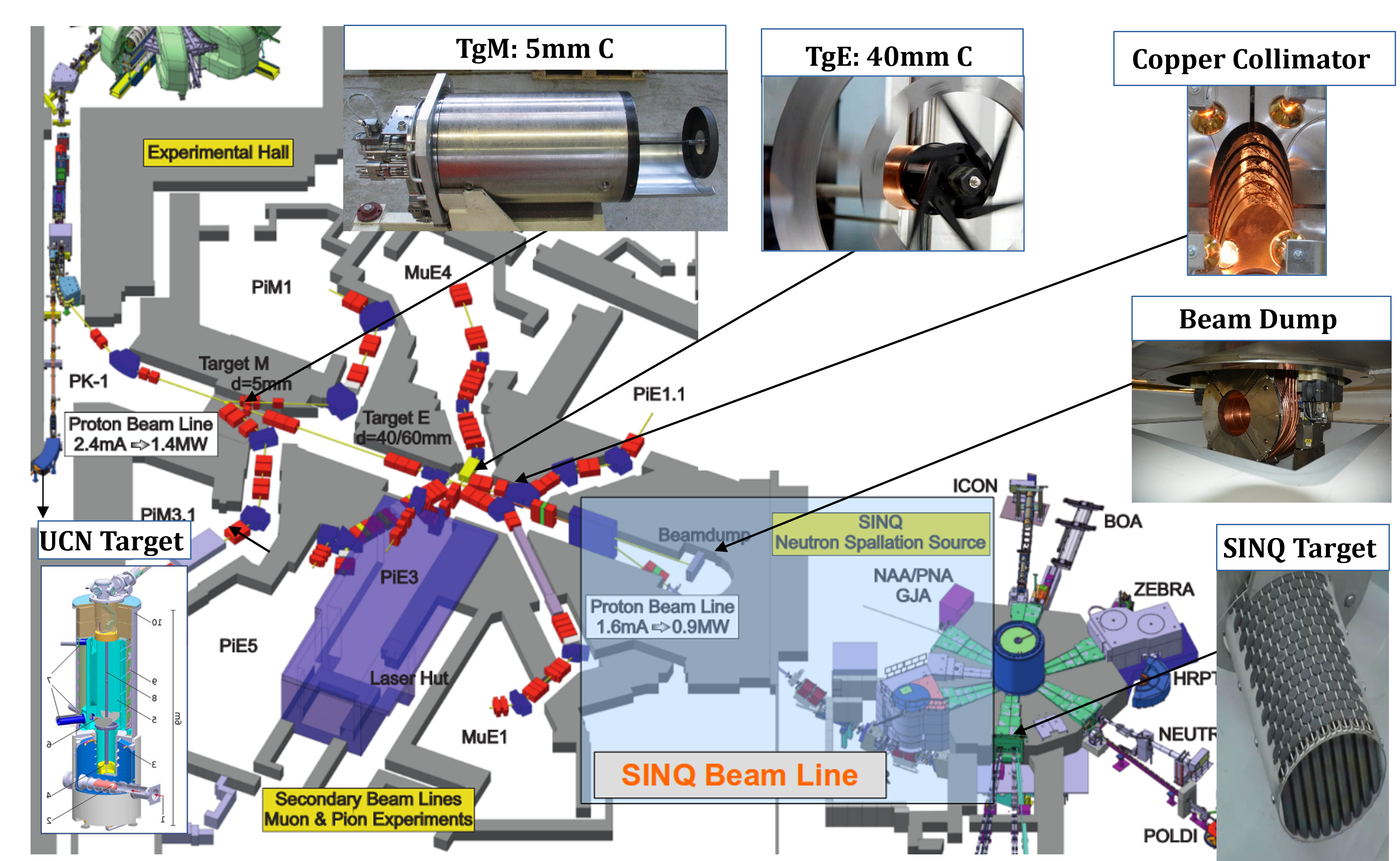}
\caption{Overview of the PSI HIPA \SI{590}{MeV} beam delivery and location of some of the most significant BIDs.}
\label{fig:hipa_layout}
\end{figure}

\section{Design and Performance of Beam Intercepting Devices at PSI-HIPA}\label{sec:present}

The HIPA proton accelerator complex currently operates the highest-power continuous proton beam worldwide, delivering up to \SI{1.4}{MW} at \SI{590}{MeV}. It supplies two in-line muon targets as well as the~SINQ (Swiss Neutron Source) and UCN (Ultra Cold Neutron source) spallation targets. High availability, around 90\%, is strongly dependent on reliable BID performance. Two rotating graphite targets, TgM and TgE, generate pions and muons. At their locations, beam power density can reach about $200$~kW/mm$^2$, creating very high temperatures, large stresses, radiation damage, and strong activation~\cite{bib:GraphiteActivity2018}. Downstream of TgE, the remaining beam is sent either to the SINQ spallation target or, when SINQ is unavailable, to a~dedicated copper beam dump. The second spallation target, UCN, is located on a different branch of the~proton channel and operated at full beam intensity thanks to a fast kicker magnet with a typical cycle of 8~s every 300~s. The \SI{590}{MeV} proton beam channel and the most relevant BIDs of the HIPA facility are depicted in Fig.~\ref{fig:hipa_layout}. A future graphite target, TgH, currently being developed within the IMPACT project (Isotope and Muon Production using Advanced Cyclotron and Target technologies), will replace TgM and provide higher-intensity muon production. In addition to these targets, the proton channel includes high-power collimators that intercept halo and scattered beam in heavily shielded regions. The~main parameters of the current and future muon production targets are summarized in Table~\ref{tab:targets}.

\begin{table}[t!]\centering
\caption{Parameters of muon production targets at PSI HIPA (present and future). TgH values are design targets under the future IMPACT program.}\label{tab:targets}
\begin{tabular}{lccc} \hline \hline
\textbf{Parameter} & \textbf{TgM} & \textbf{TgE} & \textbf{TgH} \\ \hline
Material & Graphite & Graphite & Graphite \\ 
Mean diameter~[mm] & 320 & 450 & 400 \\ 
Thickness (effective)~[mm] & 5.2 & 40 & 20 \\ 
Beam fraction intercepted (after collimation)~[\%] & $\sim$1.6 & $\sim$30 & $\sim$6 \\ 
Power deposition~[kW/mA] & 2.4 & 20 & 10 \\ 
Operating temperature (simulated)~[\textdegree C] & $\sim$850 & $\sim$1500 & $\sim$1400 \\ 
Rotation speed~[turn/s] & 1 & 1 & 1 \\ 
Design lifetime~[years] & $\sim$3 & $\sim$2 & n/a \\ \hline \hline
\end{tabular}
\end{table}

\subsection{Graphite Targets for Pion/Muon Production}

TgM and TgE are rotating graphite wheels placed sequentially in the \SI{590}{MeV} proton beam to generate pions and muons for the PSI secondary beamlines~\cite{bib:MesonTargets2021,bib:MesonTargetsSystems2021}.

\subsubsection{TgM}
TgM is the first intercepting device: a \SI{320}{mm} diameter graphite disk with \SI{5.2}{mm} effective thickness. Together with the two copper collimators located immediately downstream, it absorbs about 1.6\% of the beam. The target wheel operates near 850~\textdegree C, mainly cooled by radiation in vacuum, and rotates at 1~turn/s to spread the 2.4~kW/mA deposited heat. It is replaced every three years (with typically eight months operation per year) because of accumulated irradiation damage and wear in support components. Operationally, it has proven reliable.

\subsubsection{TgE}
TgE, located about \SI{18}{m} downstream, intercepts a much larger fraction of the beam. It is a thicker graphite wheel, \SI{450}{mm} in diameter and \SI{40}{mm} effective thickness. In operation the graphite wheel absorbs on the order of 8\% of the total beam, corresponding to roughly 20~kW/mA and around \SI{50}{kW} at full current. Steady-state graphite temperatures reach about 1500~\textdegree C. As for TgM, rotation at 1~turn/s is essential, and radiative cooling removes most of the heat. The segmented rim includes gaps to accommodate thermal expansion. Because protons missing TgE may continue to SINQ with damaging consequences, beam centering and operational reliability are critical.

\subsubsection{“Slanted” TgE Upgrade}

\begin{figure}[b]
\centering
\includegraphics[width=\linewidth]{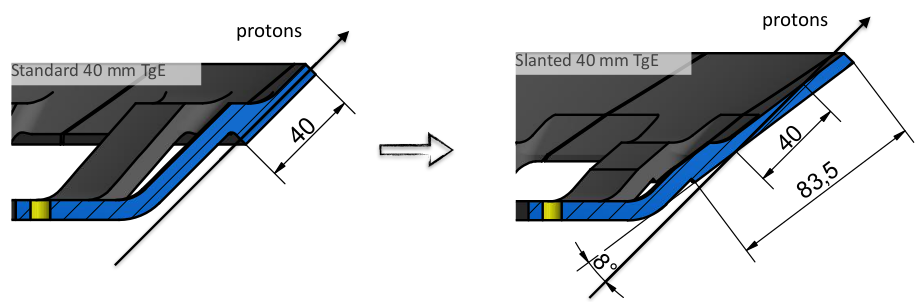}
\caption{Standard vs.\ slanted TgE geometry concept. All lengths are in mm.}
\label{fig:slanted_TgE}
\end{figure}

\begin{figure}[t]
\centering
\includegraphics[width=\linewidth]{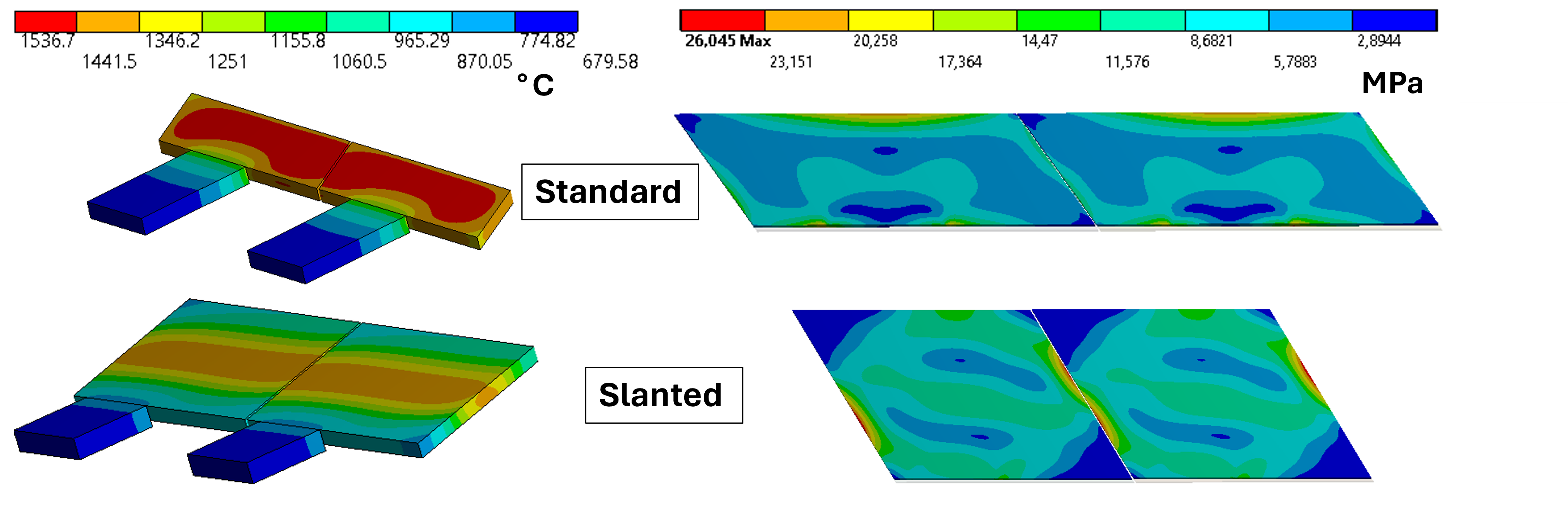}
\caption{Multiphysics (Ansys\textsuperscript{\textregistered}) simulations of temperature (left) and stress (right) distribution in two tiles of the~TgE graphite wheel for the standard and slanted geometries.}
\label{fig:targetE_sim}
\end{figure}

In 2019 PSI tested a slanted TgE geometry, finally adopted for routine operation in 2020~\cite{bib:MesonTargets2021,bib:MesonTargetsSystems2021}. Both standard and slanted concepts are shown in Fig.~\ref{fig:slanted_TgE}. By tilting the wheel by about 8\textdegree\ relative to the beam, maintaining at the same time the same effective thickness of \SI{40}{mm}, the target surface area increases significantly. This improves surface muon production and enlarges the effective target width, reducing the probability that a missteered beam bypasses the target. Both simulations and measurements have shown that the slanted concept increased usable surface-muon rates by about 50\% compared to the standard straight geometry. Thermal simulations indicated a slightly lower peak temperature for the slanted geometry compared to the standard straight design (1460 vs 1535~\textdegree C) because energy is distributed over a larger area. However, the new geometry introduces higher local stress, reaching peaks of \SI{26}{MPa}, roughly twice the stress of the straight design. Nevertheless, this larger stress value is still considered acceptable if compared to the estimated ultimate tensile stress of polycrystalline graphite at 1500~\textdegree C which is roughly \SI{38}{MPa}. Figure~\ref{fig:targetE_sim} shows simulations results of temperature and stress distribution for two tiles of the TgE wheel. To simplify the simulation process, equivalent flat geometries have been employed.

\begin{figure}[b]
\centering
\includegraphics[width=0.58\linewidth]{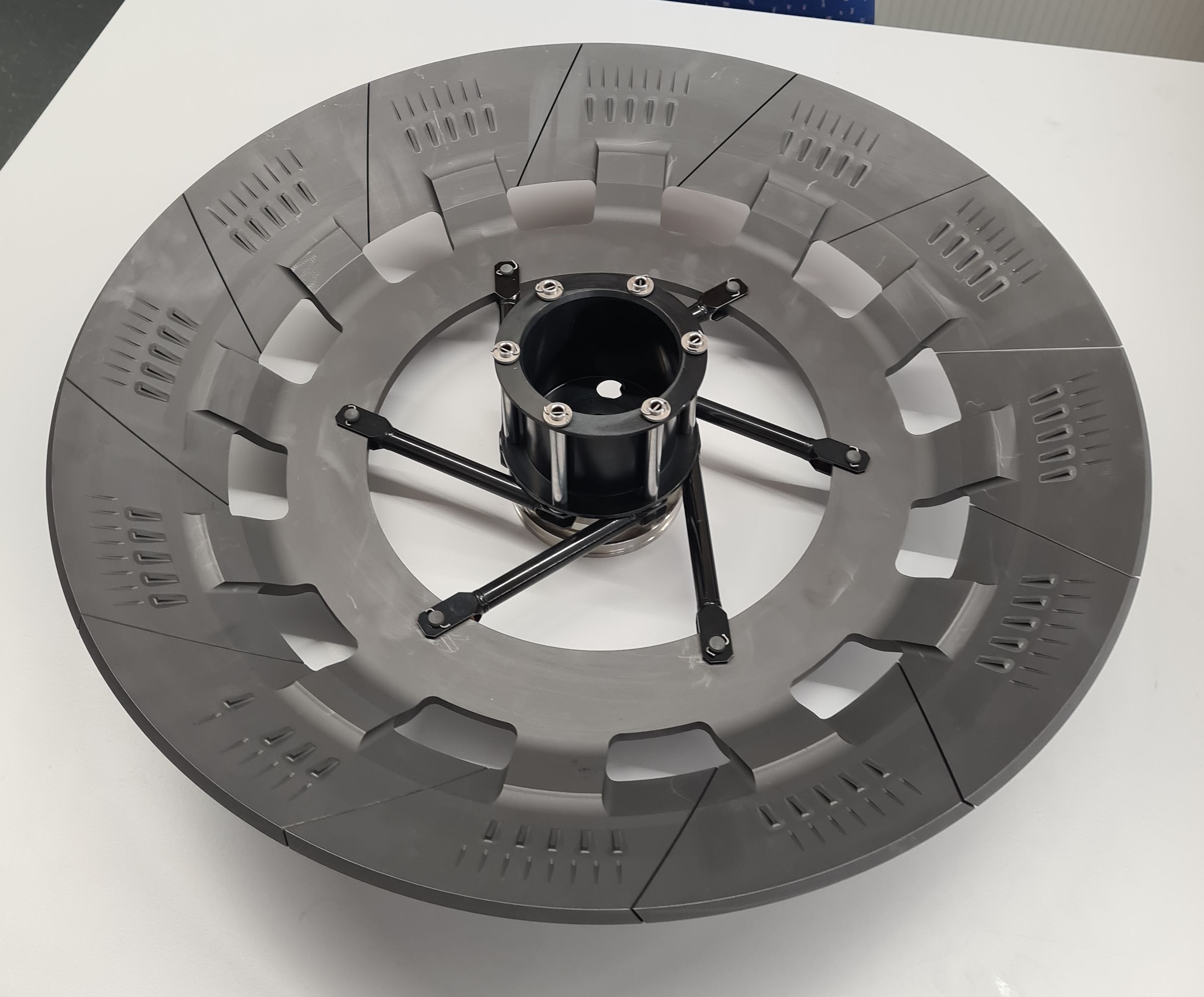}
\caption{TgE slanted graphite wheel with spokes and hub. The twelve tiles, separated by 1~mm wide gaps, can be recognised as well as the grooves and shims employed for beam position detection.}
\label{fig:targetE_wheel}
\end{figure}

To improve beam-centering diagnostics, grooves and shims were added at the target rim, as shown in Fig.~\ref{fig:targetE_wheel}. The aim was to detect when part of the proton beam misses the narrow graphite edge and continues toward SINQ. Off-center beam interception produces a characteristic modulation in the beam current at frequencies linked to grooves and shims structure, providing a possible interlock or alarm signal. The slanted target presently in operation includes these features, and data analysis is ongoing\cite{bib:SINQProtection2020}.

\subsubsection{Operational Incidents and Improvements}
\begin{figure}[t!]
\centering
\includegraphics[width=0.98\linewidth]{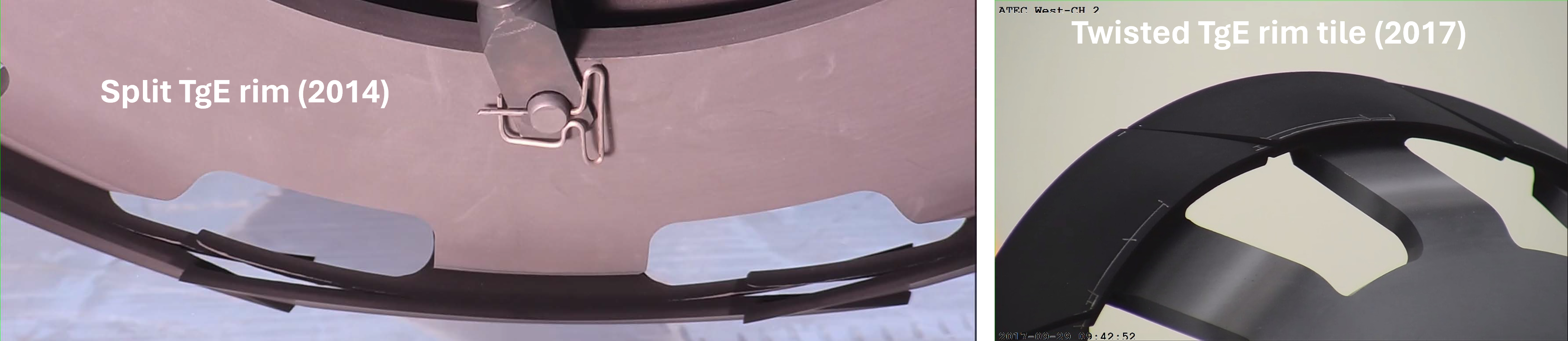}
\caption{Two examples of TgE rim failure.}
\label{fig:TgE-Fails}
\end{figure}

The reliability of the TgE has improved over the years, thanks in part to the experience gained through incidents and failures. In 2014, during a planned target exchange, it was found that the proton beam had partially cut through several graphite tiles (see Fig.~\ref{fig:TgE-Fails}, left picture). Although never fully proven, the~incident was most likely caused by a sudden drop in the TgE rotational speed during beam operation. For unknown reasons, this occurrence was not detected by the interlock system, meaning that no beam trip was triggered. This incident led to a more systematic check of parameters linked to target rotation, such as the amount of current absorbed by the target drive. In 2017, a twisted rim tile was found on the target rim. This failure had already been suspected during beam operation, since the signals of all the beam loss monitors located downstream of TgE displayed an amplitude modulation of 1~Hz, corresponding to the TgE rotation frequency. Another issue related to past TgE operation was bearing-related. Until 2021, the ceramic bearings typically lasted three to four months before failing due to heat damage. From 2022 onwards, PSI used a new bearing design developed at J-PARC, incorporating stainless steel balls and WS$_2$ lubrication blocks. Since then, TgE has operated for full annual campaigns without the need for mid-year replacement.~\cite{bib:MesonTargetsSystems2021}.

\subsection{High-Power Collimators}
Downstream of TgE, two large copper collimators, KHE2 and KHE3, intercept scattered beam and halo before SINQ~\cite{bib:MesonTargets2021}. Together they absorb about 14\% of the beam. Each device is an OFHC (Oxygen-Free High Conductivity) copper absorber with embedded stainless-steel cooling tubes. Installed around 1990, they have operated for decades under deposited beam power in the order of \SI{150}{kW}. Over time they became highly activated; in 2010, during a planned inspection, dose rates up to \SI{500}{Sv/h} were measured at the inner surface of KHE2. However, no significant damage or degradation of the copper body were observed~\cite{bib:KHE2-Inspection}.

\begin{figure}[t!]
\centering
\includegraphics[width=0.90\linewidth]{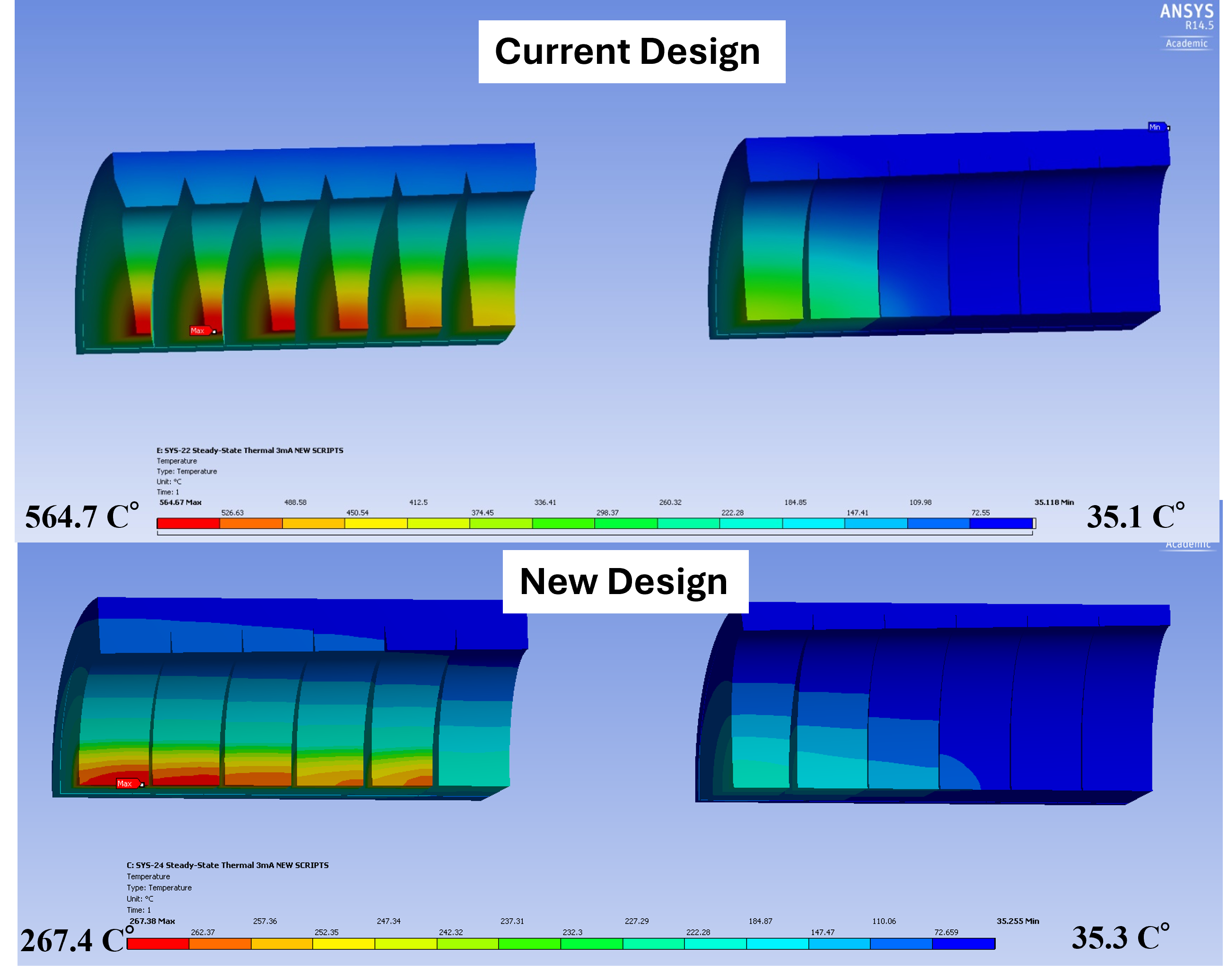}
\caption{Comparison of the simulated temperature maps for the high-power copper collimators KHE2/3 current and new design at a beam current of \SI{3.0}{mA}, corresponding to a beam power of \SI{1.8}{MW}.}
\label{fig:KHE2-sim}
\end{figure}

Planned intensity increases to \SI{3}{mA} would push the existing design beyond safe thermal limits. Simulations predicted peak copper temperatures near \SI{565}{\celsius}, well above the limit of \SI{405}{\celsius}, above which creep and phase transition significantly degrade copper performance. A new design was therefore developed with modified geometry to spread the deposited power. As shown in Fig.~\ref{fig:KHE2-sim}, the upgraded concept reduces the peak temperature to about \SI{267}{\celsius} at \SI{3}{mA}, preserving acceptable material margins.

Both the current and future collimators include narrow vertical slits in the copper blocks. These segment the absorber and allow thermal expansion, reducing constraint and stress concentration. Cooling water mass flow rate of at least 0.5~kg/s is provided in each collimator, with sufficient margin to avoid boiling. Overall, the KHE2/3 experience shows that conventional copper absorbers can operate reliably for decades, but only if cooling, stress relief, and activation management are treated as core design drivers.

\subsection{Spallation Neutron Source Target (SINQ) and Related Devices}

\begin{figure}[tb]
\centering
\includegraphics[width=\linewidth]{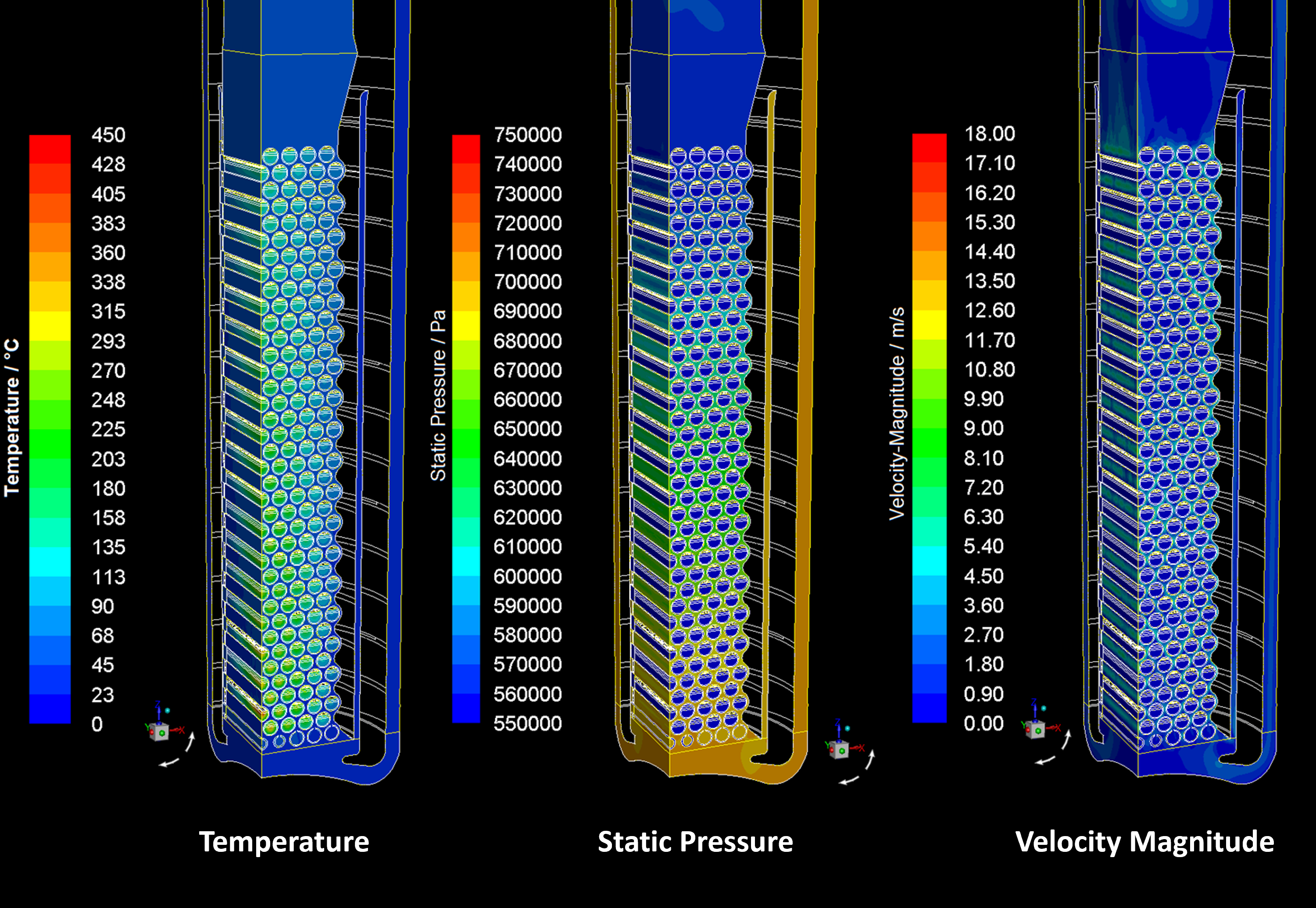}
\caption{CFD simulation results of a quarter of the SINQ target~13.}
\label{fig:SINQ-sim}
\end{figure}

Most of the remaining beam power is deposited in the SINQ spallation target downstream of the muon targets and collimators. SINQ is a heavy-water-cooled target station for neutron production~\cite{bib:SINQTarget,bib:SINQThermo}. Commissioned in 1996, the SINQ target can stop up to approximately \SI{1}{MW} of proton beam power and has a lifetime of two years. Its design has evolved through several generations to improve both neutron yield and reliability. Early solid Zircaloy rod targets (Mark I) had limited performance. The later Mark II and III 'cannelloni' concept, based on lead-filled stainless steel and zircaloy-II tubes, substantially increased neutron flux. The current Mark IV design, which has been in operation since 2009, features densely packed lead-filled Zircaloy-II tubes, as well as a lead blanket/reflector, which contribute to further improvements in neutron yield~\cite{bib:SINQMaterials2011}. In 2016, target~11 suffered a severe failure, with numerous central Zircaloy tubes rupturing and molten lead entering the cooling circuit. This led to a sudden loss of cooling and a multi-month outage. Post-irradiation analysis revealed that the damage was concentrated in the~hottest central region, where temperatures exceeded the melting point of lead. This occurrence prompted further modifications to the target design. From target~13 onwards, the central 'cannelloni', in which temperatures above the lead melting point are expected, have been replaced by full Zircaloy rods. Despite a $\approx5\%$ reduction in neutron yield, this modification, along with other improvements in proton beam line~\cite{bib:SINQProtection2020}, has greatly enhanced the safe operation of the spallation source.

The SINQ target is a prime example of the full complexity involved in a high-power BID. It must be able to absorb a significant thermal load, produce useful secondary particles, survive intense irradiation and be easily replaceable. The target development is largely based on simulation processes. Monte Carlo tools are employed to optimise the neutron yield and to map the power deposited in the~target, while CFD and coupled thermo-mechanical analyses are used to verify temperature peaks and distribution as well as ensuring adequate coolant flow in the highest-power regions. Figure~\ref{fig:SINQ-sim} illustrates a sample of CFD simulation results for target~13.

\section{Future Upgrades and New BIDs under IMPACT}\label{sec:future}
PSI’s IMPACT program foresees major upgrades starting from 2029, including new target stations and a~possible beam-current increase to \SI{3}{mA}. This requires a new generation of BIDs, most notably the TgH station and collimators for HIMB (High Intensity Muon Beams) and the TATTOOS (Targeted Alpha Tumour Therapy and Other Oncological Solutions) isotope-production target and dump~\cite{bib:IMPACTCDR2022}.

\subsection{HIMB Target Station (TgH)}

The HIMB project aims to achieve muon intensities of up to $10^{10}\,\mu^+$ / s, which is roughly two orders of magnitude higher than the present capabilities. This goal will be reached by completely redesigning the~TgM region alongside the two secondary beamlines currently in operation. The \SI{5}{mm} thick TgM will be replaced by the slanted TgH, which will increase the effective graphite thickness to \SI{20}{mm}. In order to collect the largest possible number of secondary muons, the first elements of the secondary beam line, the capture solenoids, will be located just \SI{250}{mm} from the beam-target interaction point. This constraint poses unprecedented design challenges. Furthermore, the fringe field of the capture solenoids will affect the proton beam trajectory, necessitating additional corrector magnets~\cite{bib:IMPACTCDR2022,bib:HIMB2023}.

\begin{figure}[tb]
\centering
\includegraphics[width=0.99\linewidth]{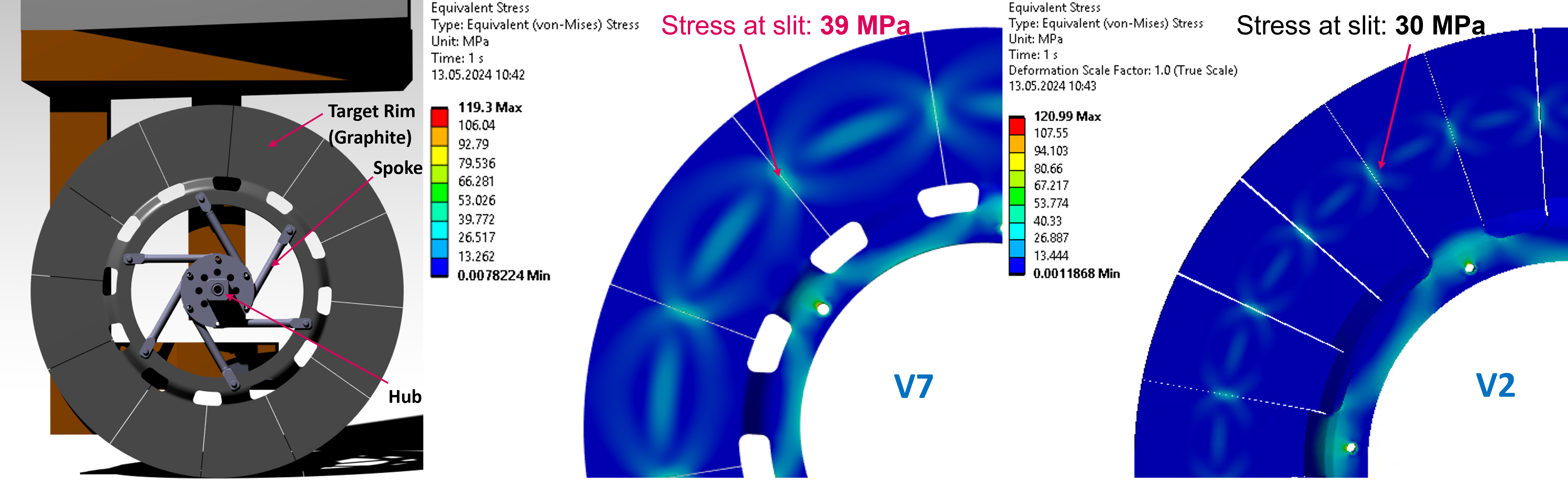}
\caption{TgH CAD 3D-model (left) and results of structural simulations of two graphite rim versions, V7 and V2.}
\label{fig:TgH-sim}
\end{figure}

The TgH design is essentially a large rotating graphite wheel with a shallow beam incidence angle of about 10\textdegree. Simulations predict peak temperatures around \SI{1400}{\celsius} at \SI{3}{mA} beam current, with total deposited power of about \SI{30}{kW}. These values are demanding but comparable to those already experienced in TgE \cite{bib:HIMB2023}. Structural integrity is the main concern. FEA shows the highest stresses at the gaps between graphite segments, where thermal gradients concentrate. Among the several considered variants, Fig.~\ref{fig:TgH-sim} shows two selected versions, V2 and V7, with 24 and 12 tiles respectively. Simulations results delivered peak equivalent stresses of roughly \SI{30}{MPa} for V2 and \SI{39}{MPa} for V7. Since the~ultimate tensile stress for graphite is about \SI{38}{MPa}, the version with the largest number of tiles is favored for the final choice, taking however at the same time factors like manufacturability, and muon yield into account~\cite{bib:IMPACTTDR2025}.

Additional protection elements are required as well: a water-cooled copper plate protects the bearings and support structure from secondary-particle heating, while an upstream Densimet\textsuperscript{\textregistered} collimator intercepts missteered beam. New downstream collimators KHH0, KHH1, and KHH2 are also planned to absorb large angle scattered particles and residual halo. Their thermal analyses indicate acceptable temperature and stress levels when stress-relief gaps and adequate cooling are included~\cite{bib:HIMB-Collimators,bib:IMPACTTDR2025}.

Overall, the TgH station represents a more integrated BID system than the present TgM: a high-load primary target surrounded by dedicated protection, cooling, and cleanup elements. It also shows how existing PSI experience is being transferred directly into future design choices.

\subsection{TATTOOS Target and Beam Dump for Isotope Production}

The TATTOOS project is intended for medical-isotope production using about \SI{100}{\micro A} of the \SI{590} {MeV} beam (corresponding to \SI{59}{kW}) in a dedicated line. Although the current is modest compared with the~main proton channel, the thermal challenge is severe because the target must run near \SI{2400}{\celsius} to enable fast isotope release. Solid tantalum is the baseline material, with UC$_x$ also considered. As \SI{26}{kW} beam power will be deposited in the \SI{200}{mm} long, \SI{100}{mm} equivalent target, even a refractory material like tantalum can approach its melting point if the power density is too high~\cite{bib:IMPACTCDR2022}.

Target development focuses on maximizing passive, mainly radiative, cooling while spreading beam power. Studied concepts include stacks of thin tantalum discs, conical bores, and beam wobbling to flatten the spatial power distribution. Simulations show how sensitive the design is to geometry and optics: in only one of the several tested configurations the target temperature stays below the tantalum melting point, but still remains above the desired operating value. Further optimization is thus required before finalization~\cite{bib:TATTOOS-Target}.

\begin{figure}[tb]
\centering
\includegraphics[width=0.99\linewidth]{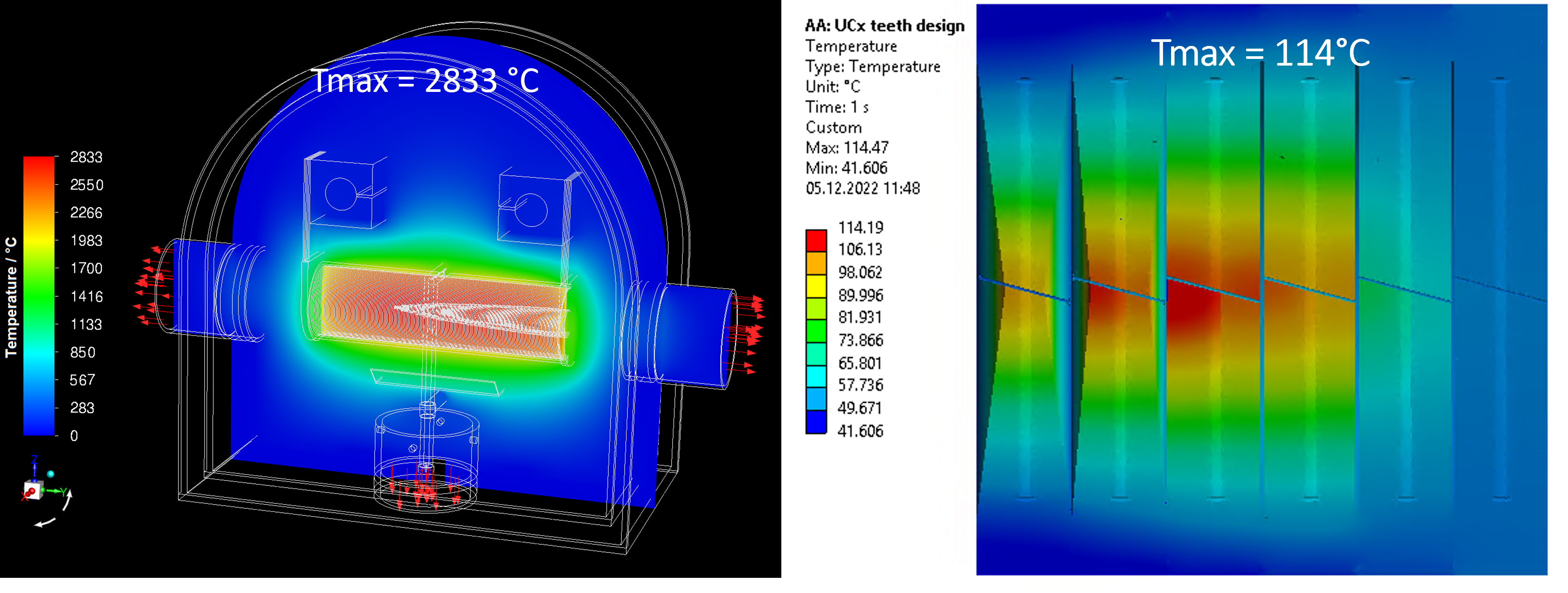}
\caption{Left: simulated temperature distribution in the TATTOOS Ta-Target for a beam current of \SI{100}{\micro A} and a~target diameter of \SI{60}{mm}. To enhance the temperature homogeneity, the beam is wobbled and the target is provided with a conical bore. The beam enters the target from the right. Right: Simulated temperature map of the~TATTOOS OFHC dump for the so-called standard scenario \SI{100}{mm} equivalent UC$_x$-Target and wobbled proton beam.}
\label{fig:TATTOOS-sim}
\end{figure}

Because the target will not absorb the full beam, the line also requires a dedicated water-cooled copper dump. Its layout follows familiar PSI design principles: OFHC copper, embedded cooling tubes, and 1.5~mm stress-relief slits. Two operating scenarios were analysed: normal operation, with the target installed and beam wobbling active, and a worst case situation in which the full beam reaches the dump. Even in the worst case peak temperature and deformation remain acceptable~\cite{bib:TATTOOS-BeamDump}.

\section{Remote Handling of Activated Components}\label{sec:remote}

\begin{figure}[b!]
\centering
\includegraphics[width=0.99\linewidth]{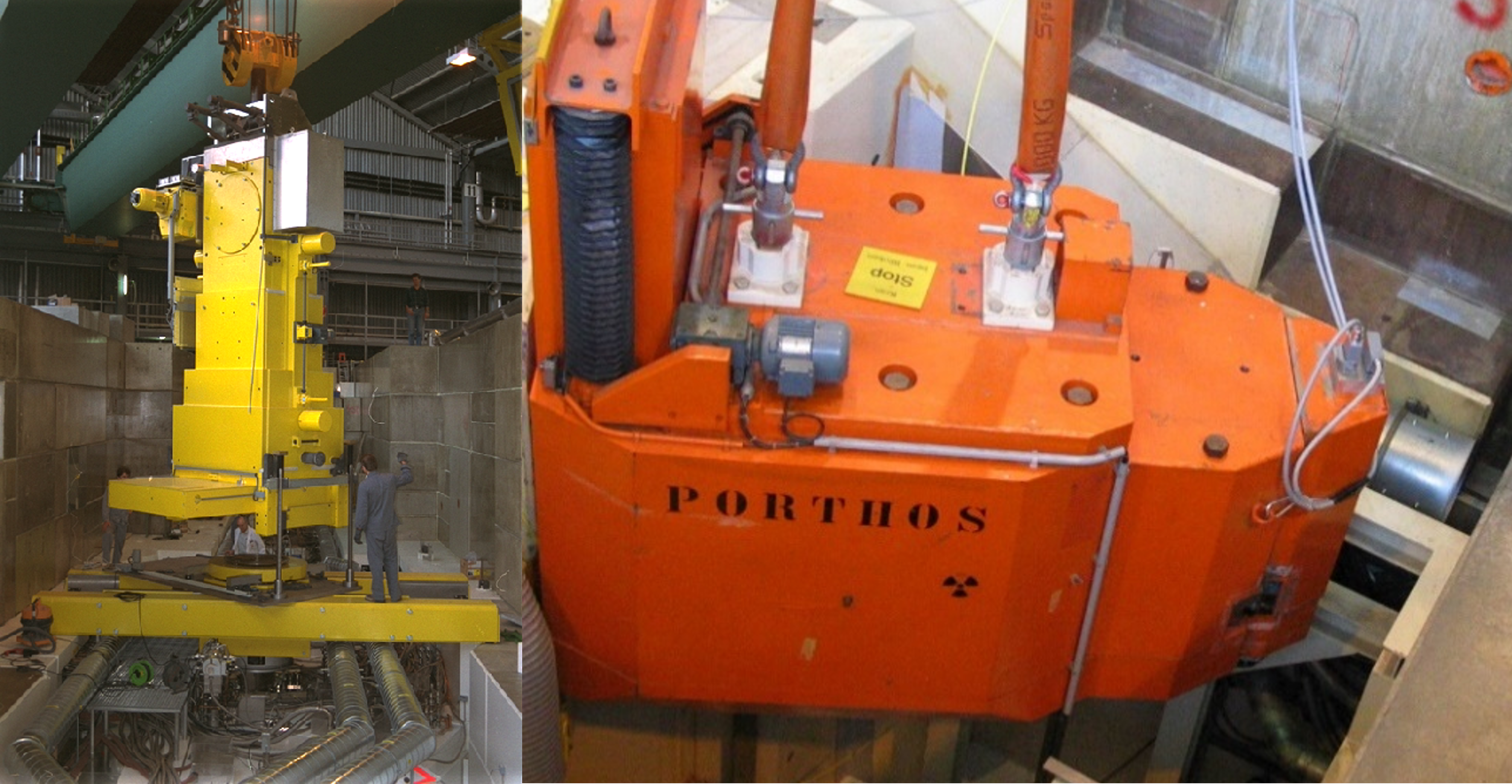}
\caption{Left: TgE exchange flask being craned to the BID location above the HIPA proton channel. The flat yellow frame under the flask is the support bridge, lying in the lateral concrete shielding blocks and used to sustain the~weight of the flask and align it to the insert of the active component to be extracted ant transported to the hot cell. Right: due to its horizontal exchange concept, the TgM flask differs from the all others employed at PSI.}
\label{fig:TgME-ExFL}
\end{figure}

All major BIDs at PSI become highly activated and must therefore be handled remotely. PSI uses dedicated exchange flasks and hot-cell infrastructure to remove, transport, and service irradiated targets, collimators, and related components. The exchange flasks are heavily shielded, remotely locked, and transferred by crane between the accelerator vault and hot-cell facilities, where manipulators can access the activated hardware. Their use has been essential in keeping personnel dose low while maintaining acceptable maintenance turnaround. Each flask is designed so that the external dose rate remains below \SI{2}{mSv/h} even with highly activated components inside. This can only be reached thanks to massive use of steel or lead shielding, which can lead to extreme heavyweight. This has to be taken into account during the design phase, for the total weight of flask and active component must not exceed the crane payload. As an example, the TgE exchange flask weighs \SI{42}{t}, leaving a maximum weight of \SI{18}{t} for the~transported component, given the crane payload of \SI{60}{t}.

Remote handling is a BID design requirement. BIDs must include remote-compatible interfaces such as defined lifting points, quick-disconnect features, alignment guides, and replaceable modular assemblies. Future installations under IMPACT are therefore being designed from the beginning for compatibility with dedicated flask systems and modular exchange concepts~\cite{bib:IMPACTCDR2022,bib:IMPACTTDR2025}. PSI’s experience shows that strong investment in remote handling improves both safety and overall facility uptime.

\section{Conclusion}
Beam intercepting devices are indispensable to high-intensity accelerator operation. At PSI HIPA they enable muon and neutron production, protect downstream systems, and make it possible to run a \SI{1.4}{MW} continuous proton beam with low uncontrolled losses. Their design, however, requires simultaneous treatment of energy deposition, heat removal, structural response, radiation damage, activation, manufacturability, and maintainability. The PSI examples reviewed here illustrate this clearly: graphite targets operate above \SI{1000}{\celsius}, copper absorbers dissipate hundreds of kilowatts, SINQ combines performance demands with robustness, and all components require remote handling.

Operational feedback is as important as simulation. Events such as TgE rim deformation and rotation issues or SINQ tube rupture have directly led to better monitoring, improved bearings and safer margins. The ongoing IMPACT program applies these lessons to a new generation of devices, including TgH and the TATTOOS target station, where slanted geometries, segmentation, advanced cooling, and dedicated protection absorbers are again central design features.

In short, BIDs remain one of the most demanding classes of accelerator components because they are where beam power is intentionally converted into heat, stress, activation, and useful secondary particles. Continued progress will depend on the same combination that has enabled PSI’s success so far: simulation, materials development, operational learning, and careful engineering integration.

\end{document}